\documentclass{aa}

\usepackage{graphicx}
\usepackage{subfigure}
\usepackage{txfonts}
\usepackage{amsmath}
\usepackage{amssymb}
\usepackage{xcolor}

\title{Growing into and out of the bouncing barrier in planetesimal formation}

\author{Maximilian Kruss
        \and Jens Teiser
        \and Gerhard Wurm}

\institute{Fakult{\"a}t f{\"u}r Physik, Universit{\"a}t Duisburg-Essen, Lotharstr. 1, 47057 Duisburg, Germany\\
        \email{maximilian.kruss@uni-due.de}}

\abstract{
In recent laboratory studies the robustness of a bouncing barrier in planetesimal formation was studied with an ensemble of pre-formed compact mm-sized aggregates. Here we show that a bouncing barrier indeed evolves self-consistently by hit-and-stick from an ensemble of smaller dust aggregates. In addition, we feed small aggregates to an ensemble of larger bouncing aggregates. The stickiness  temporarily increases, but the final number of aggregates still bouncing remains the same. However, feeding on the small particle supply, the size of the bouncing aggregates increases. This suggests that in the presence of a dust reservoir aggregates grow into but also out of a bouncing barrier at larger size. 
}


\begin{document}

\maketitle

\section{Introduction}

Sometimes research just gets stuck. This is the case for the early evolution of solids in planet formation. There is ample evidence ranging from general physics of protoplanetary disks over numerical simulations of particle collisions to laboratory experiments that small dust grows to larger aggregates in sticking collisions. This is essentially inevitable \citep{Blum2008,Testi2014}. However, it is not clear how far this growth can proceed sizewise. Combined experimental and numerical work show that particles eventually become compact and no longer stick in collisions, but start to bounce off each other \citep{Zsom2010,Guttler2010}. \citet{Zsom2010} called this the bouncing barrier. Depending on the disk settings and the particle properties the size of the bouncing particles might get as large as mm or cm, but this strongly depends on the initial conditions. \citet{Okuzumi2012} propose that ice particles might lead to the formation of sub-km sized bodies just by hit-and-stick, and that bouncing would not then be an issue.  For micrometer-sized silicates, however,  experiments by \citet{Kelling2014} and \citet{Kruss2016} showed that the existing bouncing barriers are very robust. Even though sticking efficiencies as high as 20\% might allow individual collisions to lead to sticking, the connections between compact dust aggregates are very weak and no long-term collisional growth proceeds. This poses two questions. First, do bouncing barriers also evolve in laboratory experiments if the initial condition is changed in such a way that ensembles of compact aggregates are not prepared artificially? Second, is there a way to proceed with planetesimal formation in the face of a bouncing barrier?

The latter might have different answers. \citet{Teiser2009} and \citet{Windmark2012} show that seeding the bouncing ensemble with a large body leads to its efficient growth by mass transfer collisions where the small projectile is destroyed, but adds mass to the larger target. Other solutions might exist if the size of the aggregates stuck at the bouncing barrier could be increased. Then collision velocities increase as well and might eventually also allow collisional growth again by fragmentation and reaccretion. However, \citet{Windmark2012} noted that this is not beneficial as it also includes many destructive collisions later on. Alternatively, growing in size might allow  a large particle fraction to enter a regime of Stokes numbers where streaming instability might set in and trigger gravitational collapse \citep{Chiang2010, Drazkowska2014, Johansen2014}.

We therefore continue experiments to study the bouncing barrier following the work by \citet{Kelling2014} and \citet{Kruss2016}. First, we start with a distribution of dust particles that grow by hit-and-stick collisions to see what the final state of such an ensemble is. Second, we feed small sticky dust to an existing ensemble of bouncing aggregates to see how this changes the final outcome of the collisions.

\section{Experiments}

\begin{figure}[h]
        \centering
        \includegraphics[width=0.9\columnwidth]{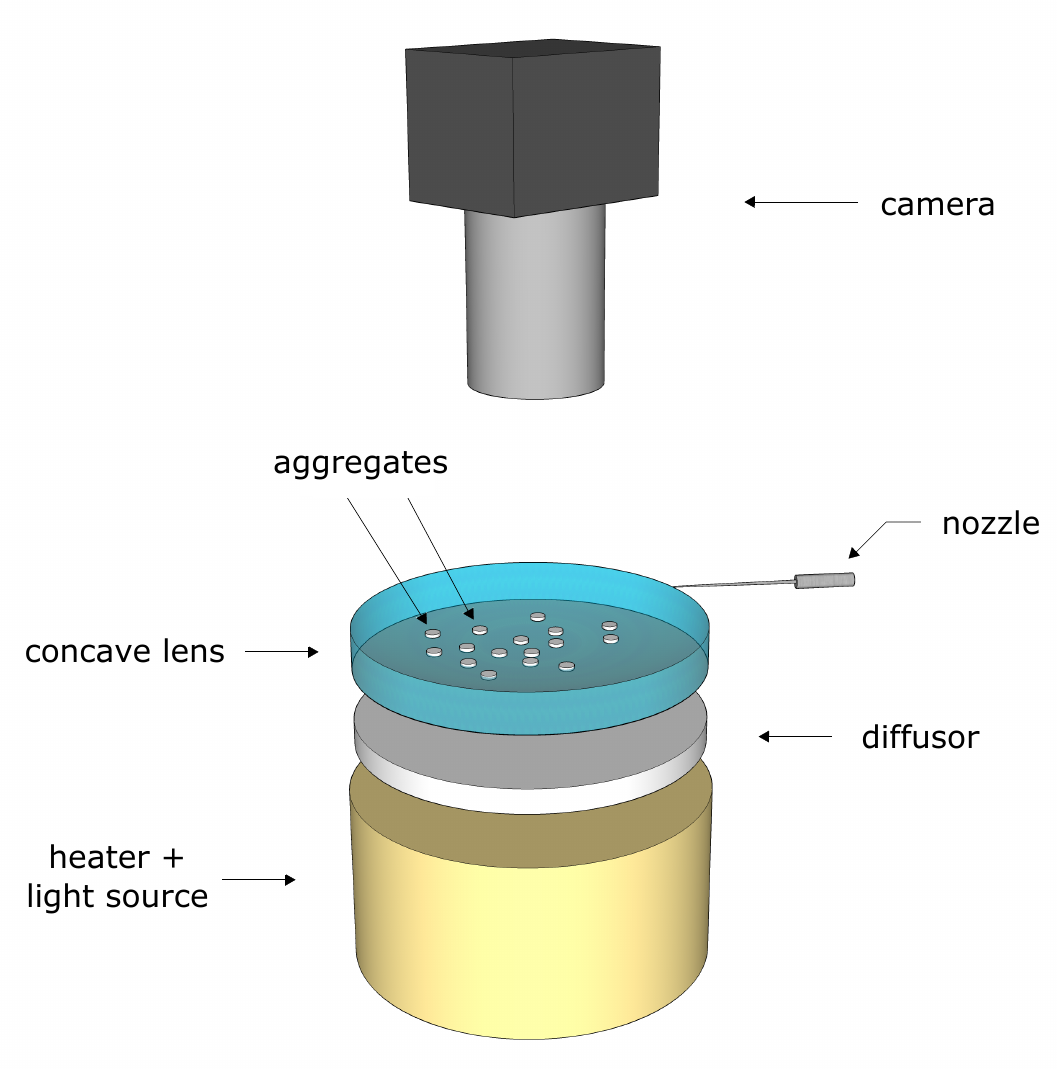}
    \caption{\label{fig.setup} Particles placed on a heater at low ambient pressure levitate, allowing  motion and collisions to be observed \citep{Kruss2016}.}
\end{figure}

The experimental setup is shown in Fig. \ref{fig.setup}. It is the same as described in \citet{Kruss2016} improving earlier work by \citet{Kelling2014}.  In summary, it works in the following way. Dust is placed on a heater at around 900 K. If the ambient pressure is reduced to the millibar range, gas flows from the cooler top through the pores of dust aggregates towards the hotter bottom (a procedure known as thermal creep). This leads to an overpressure that lifts the dust, which is now free to move around in 2D, collide, and grow or bounce.

The advantage is that a continous and self-consistent evolution of a freely moving dust sample can be observed. Also, the velocity distribution is not damped over time. This mimics the conditions in protoplanetary disks very well as velocities between collisions are reset by coupling to the surrounding gas. This is in contrast to granular gases, which can cool collisionally, which usually leads to a clumping of particles according to Haff's law \citep{Haff1983,Luding1999,Harth2016}. The disadvantage is that a levitation technique is still only a levitation and not 3D free motion. Aggregates and clusters of aggregates can only move randomly in 2D. Gravity is still present and the leaking supporting gas cushion leads to gas motions. While these random outflows generate relative velocities, it comes with residual forces acting during collisions. However, in earlier work these effects were estimated to be negligible in the context of these experiments \citep{Kelling2014}.
 
\section{Evolution of small dust towards bouncing}

\begin{figure}[h]
\includegraphics[width=\columnwidth]{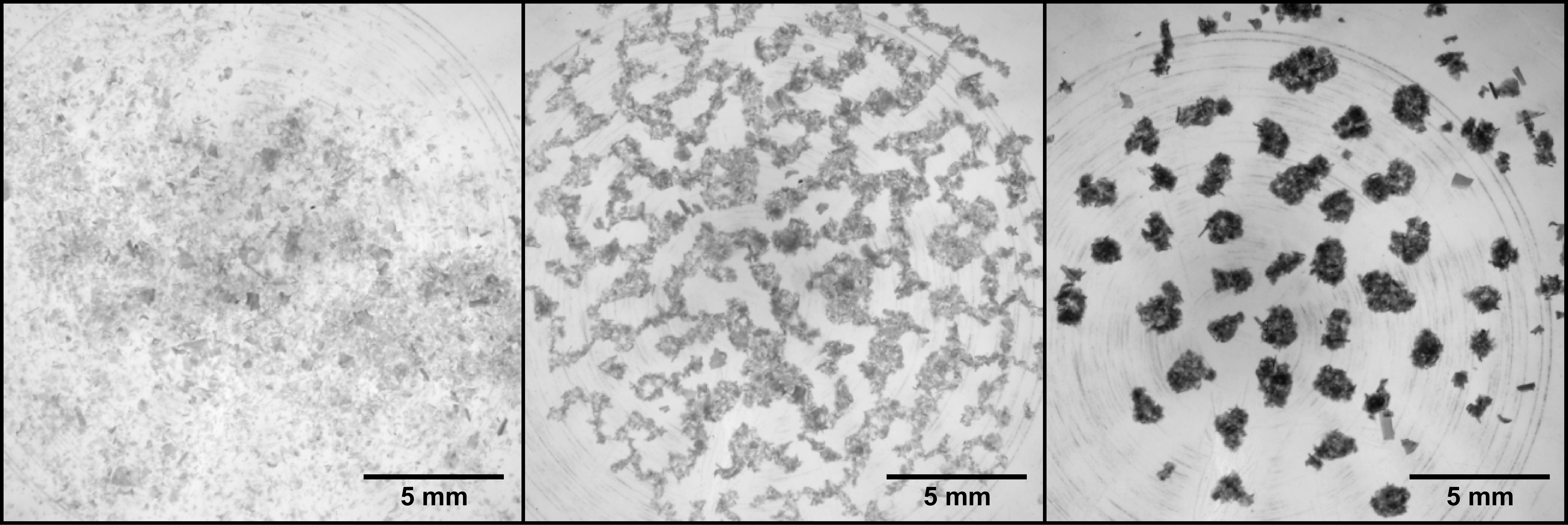}
    \caption{\label{fig.dustevolution} Left: Quartz dust is placed on the heater; Middle: Larger entities grow after several minutes; Right: At the final stage a constant number of large aggregates form, which only bounce off each other in the long term.}
\end{figure}

We covered the surface of the experiment with $\rm SiO_2$ (quartz) dust consisting of micrometer grains -- again as in the earlier work \citep{Kelling2014, Kruss2016}. This is seen in Fig. \ref{fig.dustevolution} (left). We did not prepare the dust in a special way. The overall density of the sample is low, which results in low contrast. On the other hand, the number density is large. Owing to the small size of the units and the low contrast, we did not attempt to determine a velocity distribution of individual aggregates.
 
In Fig. \ref{fig.dustevolution} (middle) an intermediate state of the dust evolution is shown after several minutes. The spatial distribution shows that the dust is concentrated to a certain degree leaving a large unfilled space and more dense dust-filled areas. This increases the contrast. However, also here, collision velocities are not well defined as the aggregates are not yet fully compacted, and parts of individual structures move in different ways as they are not strongly connected to each other. Figure \ref{fig.dustevolution} (right) shows the final formation of individual aggregates of similar size, which are more dense now and are no longer concentrated  by agglomeration. These aggregates in our setup have absolute and relative velocities on the order of 10 mm/s on average.

\begin{figure}[h]
\includegraphics[width=\columnwidth]{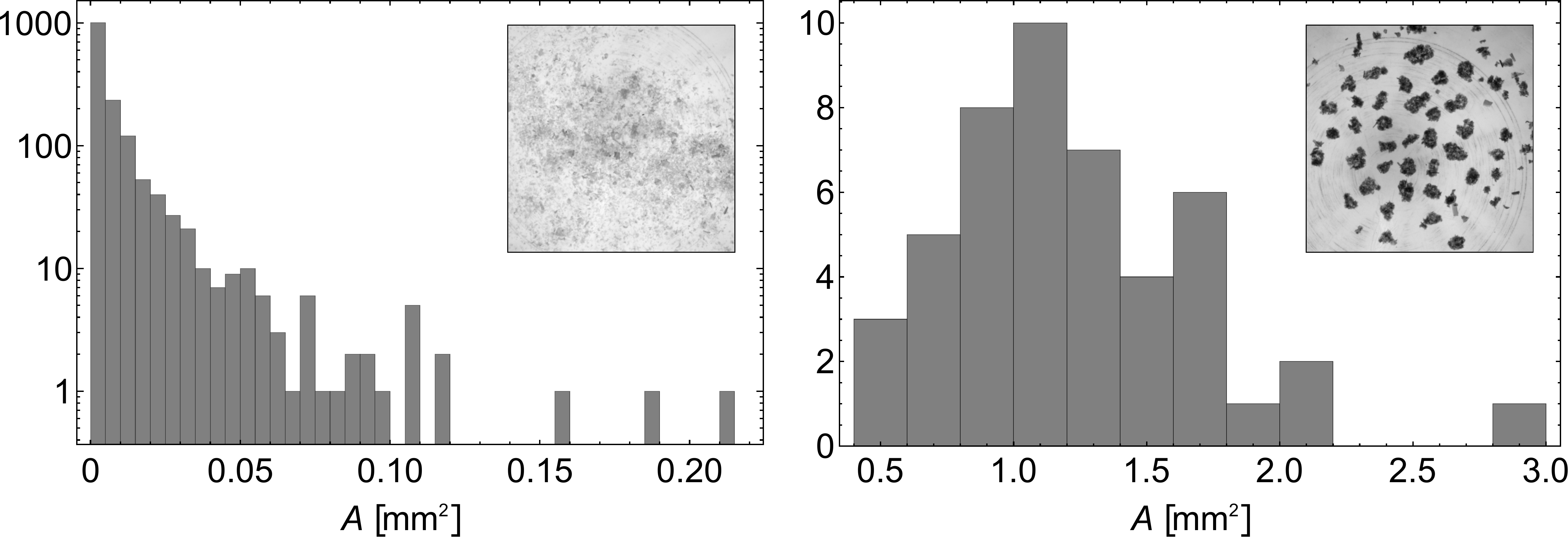}
    \caption{\label{fig.sizetime} Size distribution of connected structures for the initial configuration and a late distribution. From a broad size distribution of small particles, an aggregate distribution at the bouncing barrier evolves.}
\end{figure}

The size of connected structures in the images can be quantified; an example  is shown  in Fig. \ref{fig.sizetime} where the connected areas $A$ of structures below a certain brightness threshold are shown for two times. While the initial dust aggregate size distribution follows a power law on the order of $A[\text{mm}^2]^{-1.6}$, the further aggregation leads to the formation of larger particles distributed around 1~mm$^2$. As the smaller ones, which are located in the outer parts of the platform, do not take part in the interaction, Fig. \ref{fig.sizetime} (right) only includes aggregates with $A>0.4$~mm$^2$. The final porosity of these aggregates is $0.79 \pm 0.02$ in our case.

In any case, the experiments show that a distribution of dust in slow motion evolves naturally towards a bouncing barrier as suggested by \citet{Zsom2010} based on individual experiments and simulations \citep{Guttler2010}.

\section{Feeding small porous dust}

So far the results of \citet{Zsom2010}, \citet{Kelling2014}, \citet{Kruss2016}, and this study prove the existence of a solid bouncing barrier. Methods for overcoming this in collisions of non-icy dust suggested so far include the existence of a larger seed \citep{Teiser2009, Windmark2012}, but this still keeps the bouncing barrier as the fundamental element. Growing by instabilities likely requires cm-sized aggregates corresponding to certain Stokes numbers. If the bouncing aggregates are too small these Stokes numbers might not be reached \citep{Bai2010, Drazkowska2014}. It is therefore a question of what the size of the aggregates at the bouncing barrier actually is, and whether it can be changed. It certainly depends on the initial dust distribution, grain sizes, relative velocities, and so on. It might therefore be argued that there are   special conditions favorable for large aggregate formation (see the extremes for submicron water ice in \citealt{Okuzumi2012}).

\begin{figure}[h]
        \includegraphics[width=\columnwidth]{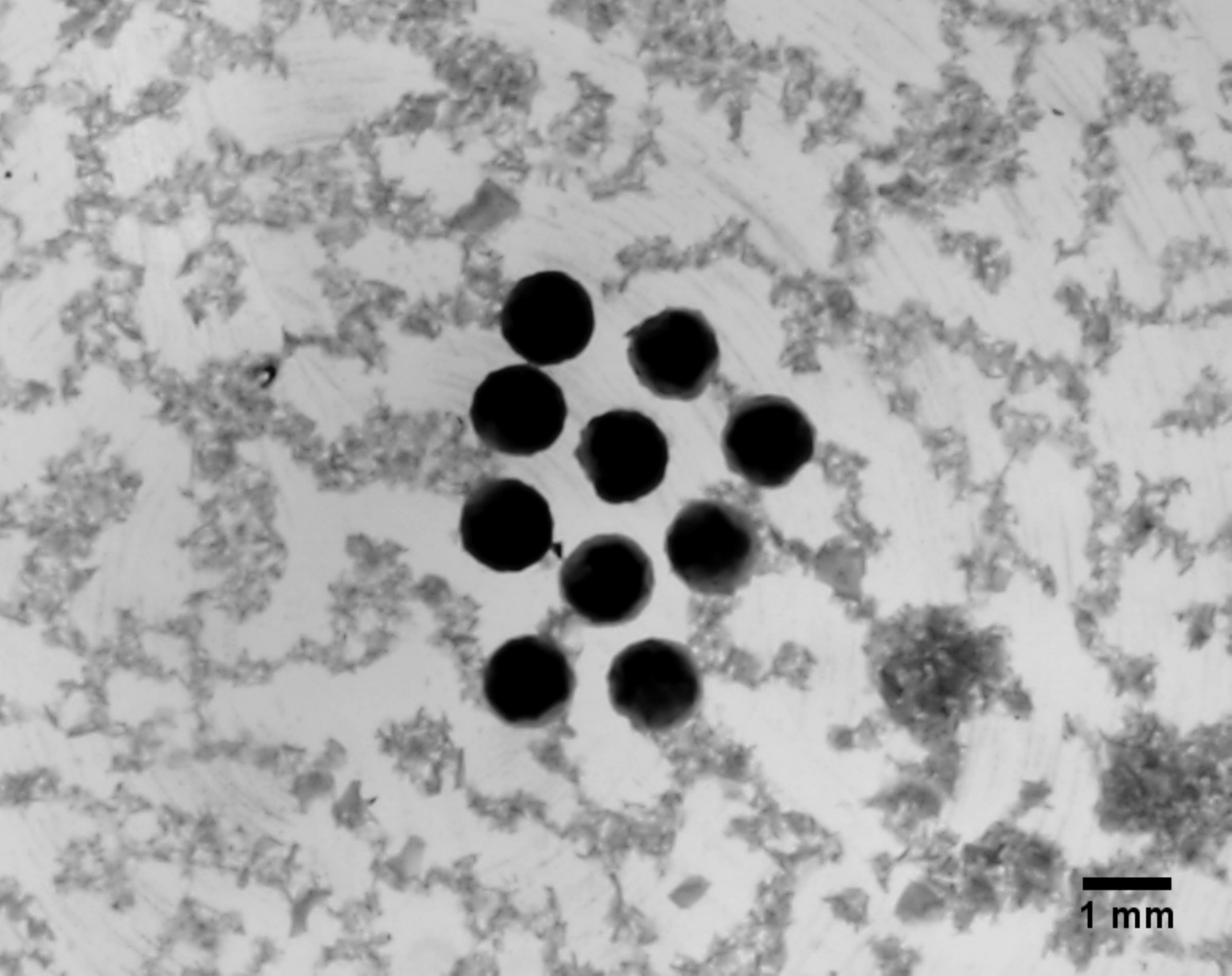}
    \caption{\label{fig.initial} Small dust aggregates are added to compact aggregates (intermediate time corresponding to Fig. \ref{fig.dustevolution} middle).}
\end{figure}

However, there might be a mechanism to increase the size of the bouncing barrier particles systematically in any case. We consider a situation where there is  a reservoir of smaller aggregates. It might be provided by the grinding of already existing larger bodies, for example that drift inwards from farther out where formation would be possible. Another potential source of smaller aggregates might be dust set free by sublimation of dust/water aggregates crossing the snowline \citep{Saito2011}. We do not specify this source  here. The question is whether such a small particle reservoir can serve as a glue that  leads to stable clustering of larger aggregates, as we expected. Figure \ref{fig.initial} shows the initial configuration where we added dust to an ensemble of compact aggregates. As before, the dust is not treated in a special way. It quickly starts to aggregate and build porous structures. This can also be observed in the evolution shown in Fig. \ref{fig.dustevolution}. The low contrast implies a low porosity, which goes with the ability to dissipate energy in collisions by restructuring. Otherwise we do not consider the exact porosity or structure to be important in this context, yet. The bouncing aggregates were formed as in the earlier studies by \citet{Kruss2016} by filling dust in molds and compressing this to filling factors of about 0.33.

\begin{figure}[h]
\includegraphics[width=\columnwidth]{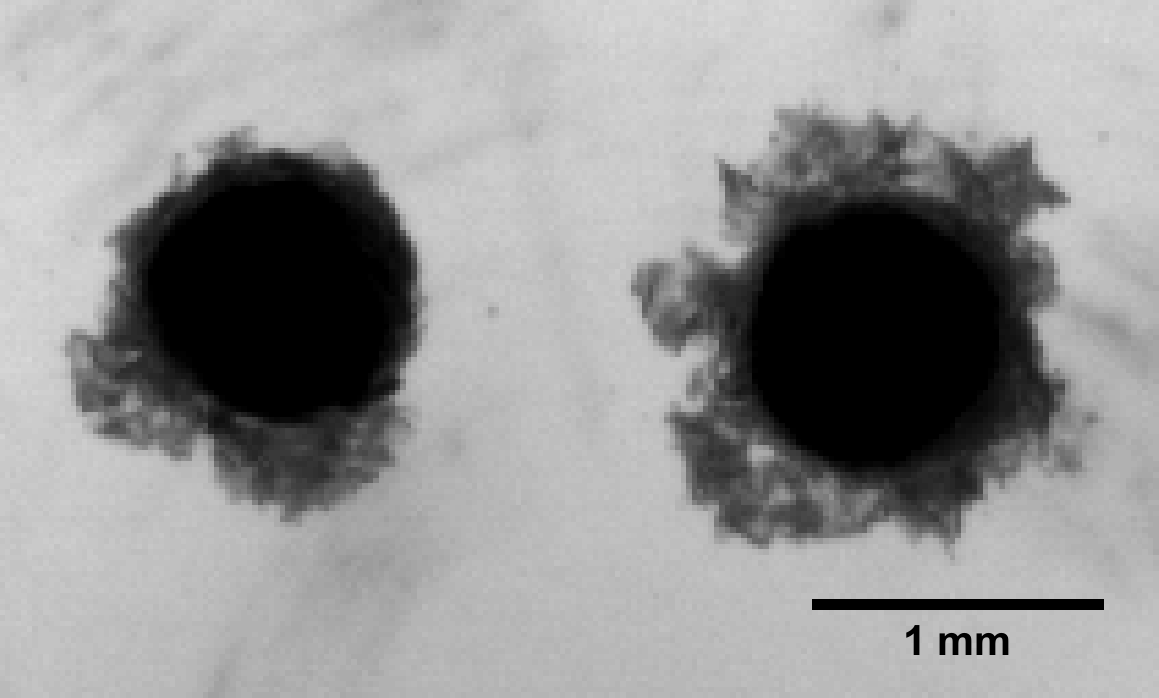}
    \caption{\label{fig.stick1}  Small dust  adheres easily to the large aggregates and enhances the sticking between large aggregates.}
\end{figure}

\begin{figure}[h]
\includegraphics[width=\columnwidth]{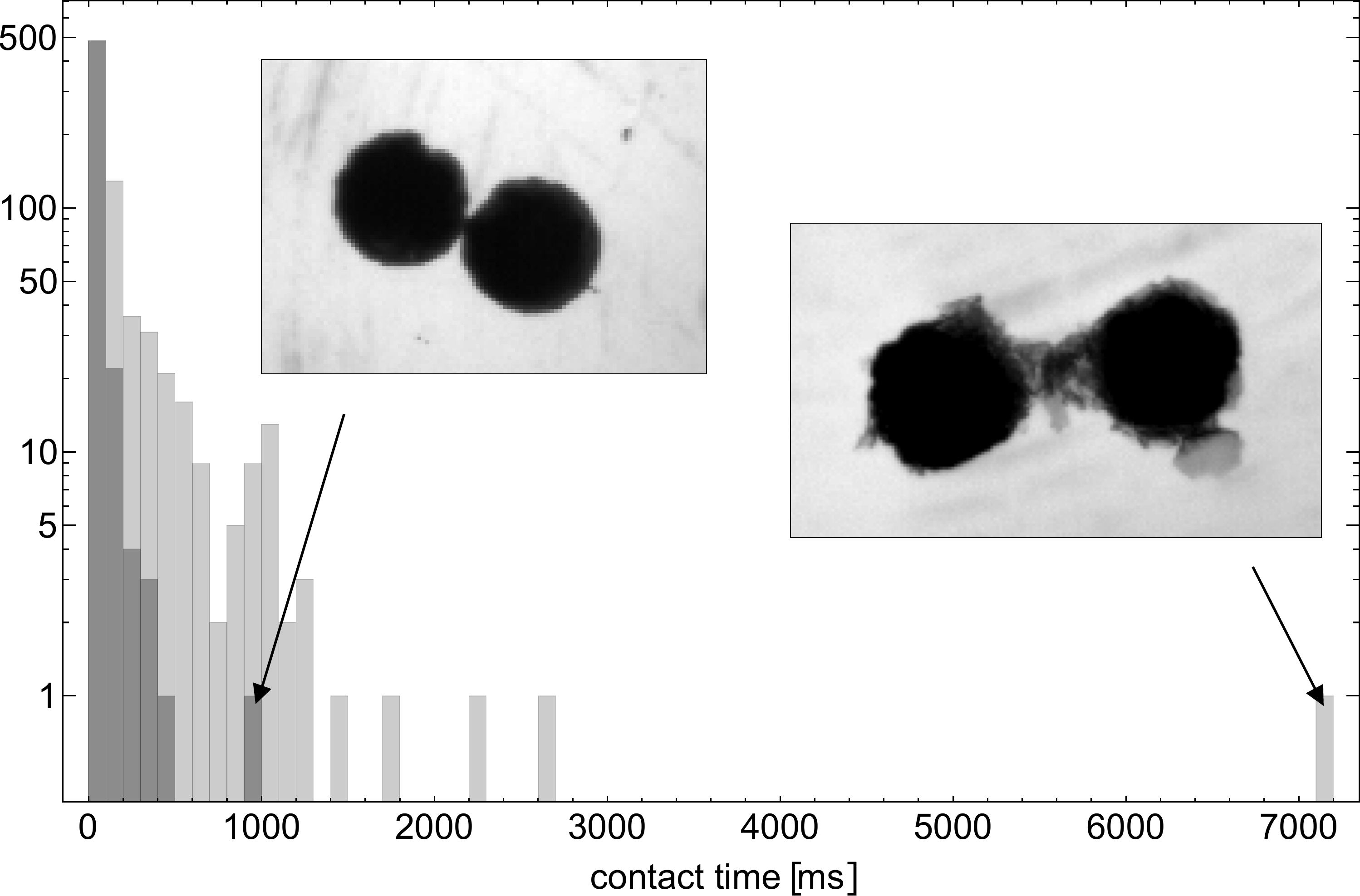}
    \caption{\label{fig.stick2}If small dust is present (light gray) the aggregates stay in touch with each other for a longer time, due to enhanced sticking compared to the case without any small dust (dark gray).}
\end{figure}

\begin{figure}[h]
\includegraphics[width=\columnwidth]{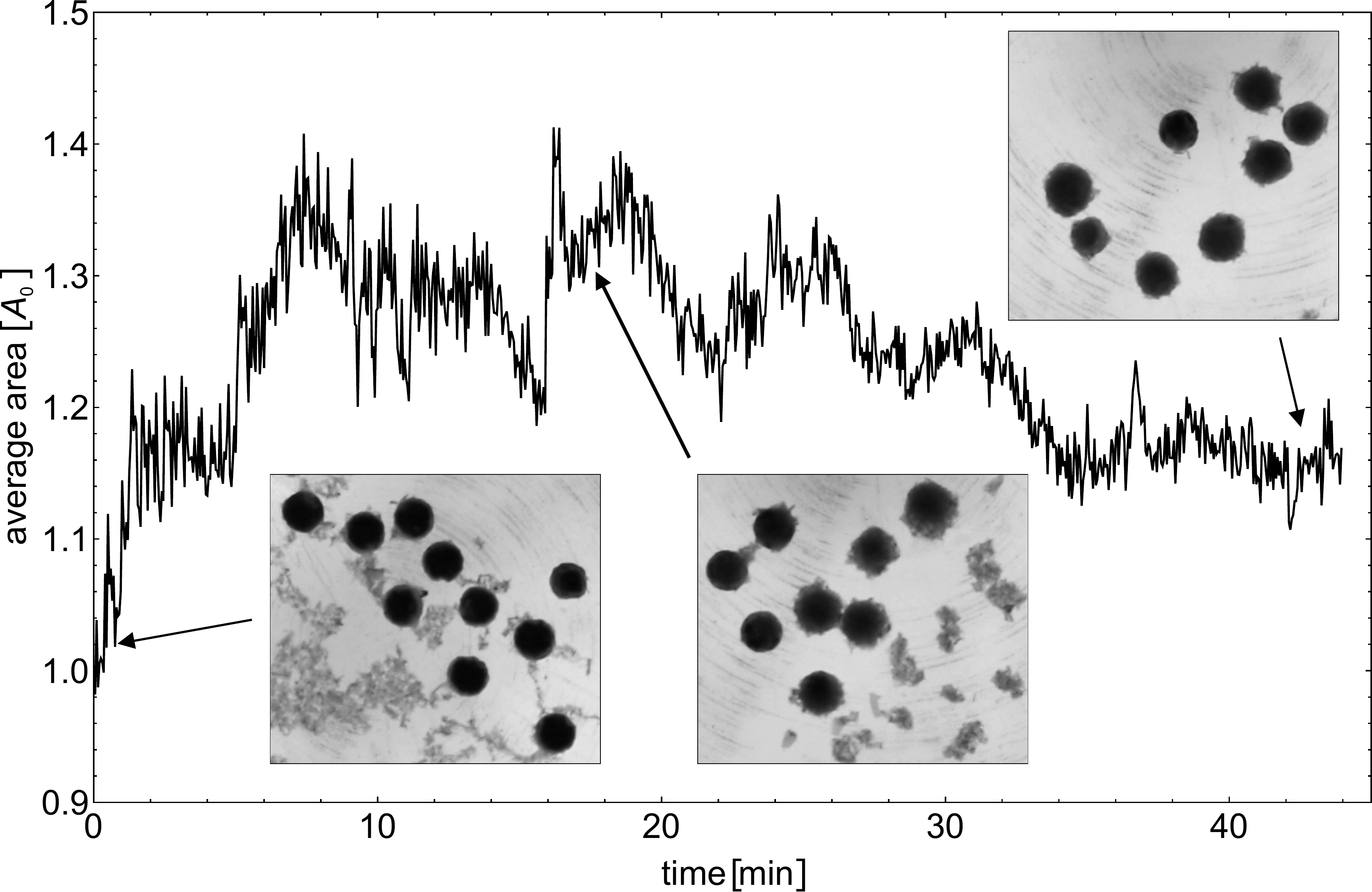}
    \caption{\label{fig.evolveme} Evolution of the average aggregate area/size over time normalized to the size $A_0$ of the initial average bouncing aggregates.}
\end{figure}

Indeed the initial evolution follows the expectations. The small dust sticks efficiently to the preformed larger aggregates (Fig. \ref{fig.stick1}). This also leads to more efficient sticking between the larger aggregates (Fig. \ref{fig.stick2} inset). Figure \ref{fig.stick2} also shows that the glued aggregates are in contact over longer times. However, the contact times are still limited and as evolution proceeds, the experiments show that this does not prevent detaching collisions at the given velocities (of up to 40 mm/s; see \citealt{Kruss2016}). The rims of fresh dust are compacted and this results in  the same number of large aggregates  as were initially  placed in the experiments. Evidently, bouncing barriers are also insensitive to particle feed since the small dust is just gathered by the larger aggregates and does not induce the formation of stable clusters. This might nevertheless be beneficial for planetesimal formation. Figure \ref{fig.evolveme} shows that the size of the aggregates grows as they share the small dust provided. Therefore, the net effect of a small particle reservoir is to continuously increase the size of the aggregates at the bouncing barrier. 

Independent of any initial size variations due to individual disk settings, one way to increase the size of mm-sized aggregates is the interaction with smaller particles. In the experiment this is realized by adding a population of small dust aggregates that can subsequently be gathered by the larger aggregates (Fig. \ref{fig.evolveme}). Applied to planet formation this might slowly grow particles into a regime where instabilities or mass transfer might take over and therefore grow particles out of the bouncing barrier.

\section{Caveat}

The aggregates are subject to the residual forces and restrictions of the levitation mechanism mentioned above, and this might not be exactly the same as having free collisions of smaller aggregates. In this sense this is not a perfect experiment to study free collisions of aggregates. However, this is the closest that might currently be done to approach pre-bouncing evolution on the ground. We argue that it is not critical for the compaction and growth if two more compact aggregates are connected by a weaker bond which might be disconnected during evolution. 

Microgravity experiments might be more suitable. For example, self-consistent growth of fractal dust was studied by \citet{Blum2000}. Further experiments aiming at larger aggregates are currently being  planned for longer duration microgravity \citep{Brisset2016}. So far the results presented here should  be considered as a strong indication for a possible way to overcome the bouncing barrier rather than a definite proof. Its application to protoplanetary disks depends on a large number of parameters including an as  yet unknown mechanism to ensure the supply with small dust particles.

The mass ratio between porous and compact dust was not specified. However, this is not critical here. If the small dust fraction dominated -- the extreme would be only small dust -- we would end up with bouncing grains as shown before.  With moderate feeding as studied here, the number of bouncing aggregates would not change. Therefore, it is unlikely that a fine tuning of the ratio between small porous and compact large particles could lead to the formation of a large body right away. It will always end up with a bouncing ensemble, only the size of the bouncers is not fixed.

\section{Conclusion}

Bouncing barriers form as a certain step in planet formation. It seems likely that particles grow into a bouncing barrier unless specific conditions exist \citep{Okuzumi2012}. However, the size of the composing aggregates is not fixed. Interestingly, the size can change if more small, porous dust is provided as this dust does not just form new bouncing particles, but attaches itself to existing bouncing particles. The exact evolution of this process depends on the mass ratio between small dust and larger aggregates in real disk settings. The number of large aggregates stays constant but they increase in size only if the existing large aggregates can pick up the particle supply before it grows into large aggregates. However, if the ratio is in the right range the bouncing barrier might continuously shift towards larger sizes until instability mechanisms  take over. If this is efficient enough, aggregates might not just grow into but also out of the bouncing barrier.

\section{Acknowledgements}

This work is funded by the DFG as KE 1897/1-1. We thank C. P. Dullemond for reviewing this paper.

\bibliographystyle{aa}

\bibliography{biblio}

\end{document}